\documentclass[final,5p,times,twocolumn]{elsarticle}
\biboptions{numbers,sort&compress}
\usepackage{amssymb}
\usepackage{lipsum}
\usepackage{comment}
\usepackage{amsmath}
\usepackage{subfig}
\usepackage[american]{babel}

\usepackage{tikz}
\usepackage{hyperref}

\journal{High Energy Astrophysics}

\makeatother
\begin{document}

\begin{frontmatter}

\title{The Study of a Cosmic Ray Candidate Detected by the Askaryan Radio Array}

\author[]{Shoukat Ali, Dave Z. Besson}
\affiliation{organization={Department of Physics and Astronomy, The University of Kansas},
            addressline={}, 
            city={Lawrence},
            postcode={66045}, 
            state={KS},
            country={USA}}
\author{(ARA Collaboration)}
\affiliation{Presented at the 32nd International Symposium on Lepton Photon Interactions at High Energies, Madison, Wisconsin, USA, August 25-29, 2025}
\begin{abstract}
Experiments designed to detect ultra-high energy (UHE) neutrinos using radio techniques are also capable of detecting the radio signals from cosmic-ray (CR) induced air showers. These CR signals are important both as a background and as a tool for calibrating the detector. The Askaryan Radio Array (ARA), a radio detector array, is designed to detect UHE neutrinos. The array currently comprises five independent stations, each instrumented with antennas deployed at depths of up to 200 meters within the ice at the South Pole.

In this study, we focus on a candidate event recorded by ARA Station 2 (A2) that shows features consistent with a downward-going CR-induced air shower. This includes distinctive double-pulse signals in multiple channels, interpreted as geomagnetic and Askaryan radio emissions arriving at the antennas in sequence. To investigate this event, we use detailed simulations that combine a modern ice-impacting CR shower simulation framework, FAERIE, with a realistic detector simulation package, AraSim.
We will present results for an optimization of the event topology, identified through simulated CR showers, comparing the vertex reconstruction of both the geomagnetic and Askaryan signals of the event, as well as the observed time delays between the two signals in each antenna.
 
\end{abstract}


\end{frontmatter}




\section{Introduction}\label{intro}
Radio detection has become a powerful technique for studying UHE CRs and neutrinos. Experiments such as the Pierre Auger Observatory and LOFAR have successfully demonstrated this approach by observing radio emissions produced by CR-induced air showers~[\cite{pa, lofar}]. Radio neutrino detectors, such as ARA, are also sensitive to these emissions, which can constitute an important background for neutrino searches~[\cite{deVries2016}]. At the same time, these signals serve as a valuable tool for detector calibration and for improving overall detection performance.

CR–induced air showers produce two types of radio emission, known as geomagnetic and Askaryan emission. The geomagnetic mechanism dominates in the atmosphere, where charged particles are deflected by Earth’s magnetic field, generating coherent radio waves. When part of the shower penetrates the ice, it produces a strong Askaryan emission through the coherent Cherenkov process, resulting from the net negative charge excess in the cascade moving faster than the speed of light in the medium~[\cite{Schroeder2025}].

ARA is primarily designed to detect Askaryan emissions from UHE-neutrino-induced showers in the Antarctic ice~[\cite{allison_ara}]. Currently, five stations, including one phased array, are deployed at the South Pole. A typical ARA station consists of instrumentation in four boreholes, known as strings. Each string contains two pairs of receiving antennas, with one antenna in each pair sensitive to horizontally polarized (HPol) signals and the other to vertically polarized (VPol) signals. The antenna frequency response extends from 130 to 850 MHz. Additionally, two strings are dedicated to calibration pulsers, each equipped with a pair of HPol and VPol transmitting antennas used for detector calibration~[\cite{ara}]. 

An event exhibiting a double-pulse structure (two distinct peaks observed across multiple channels) was recorded by A2 and later identified in a neutrino search analysis~[\cite{brian}]. This event passed all background rejection criteria for thermal and anthropogenic sources applied to the full dataset, covering eight station-years. It was initially reported as a CR candidate at APS conferences~[\cite{ammy_aps, ali_aps}] and a detailed study of the event was recently published in Proceedings of Science ~[\cite{a2_event_icrc}]. An independent search analysis was also performed to estimate double-pulse events in the 2015–2016 dataset. Only this event and a calibration pulser event passed the search. The calibration pulser event has a reconstructed vertex consistent with the known locations of the calibration pulsers, suggesting that this background event likely resulted from two pulsers transmitted simultaneously~[\cite{a2_event_icrc}].

This study presents preliminary results for a proposed event topology, comparing the vertex reconstruction of the two observed pulses with the expected geomagnetic and Askaryan emissions from an inclined CR-induced air shower. The first pulse is attributed to the in-air geomagnetic emission, while the second corresponds to the in-ice Askaryan emission, as the geomagnetic signal is expected to reach the antennas slightly earlier than the Askaryan signal. This analysis also compares the results for the measured time delays between the two pulses in each channel to the differences in expected arrival times of the two types of emission from the CR-induced air shower.\\
To characterize the polarization of the signals, the power ratio recorded by HPol and VPol for the two pulses was estimated and compared with that expected from the geomagnetic emission relative to the Askaryan emission.
\section{Event Features}\label{ara_det}
\begin{figure}[hbt!]
    \centering
    \includegraphics[width=0.5\textwidth]{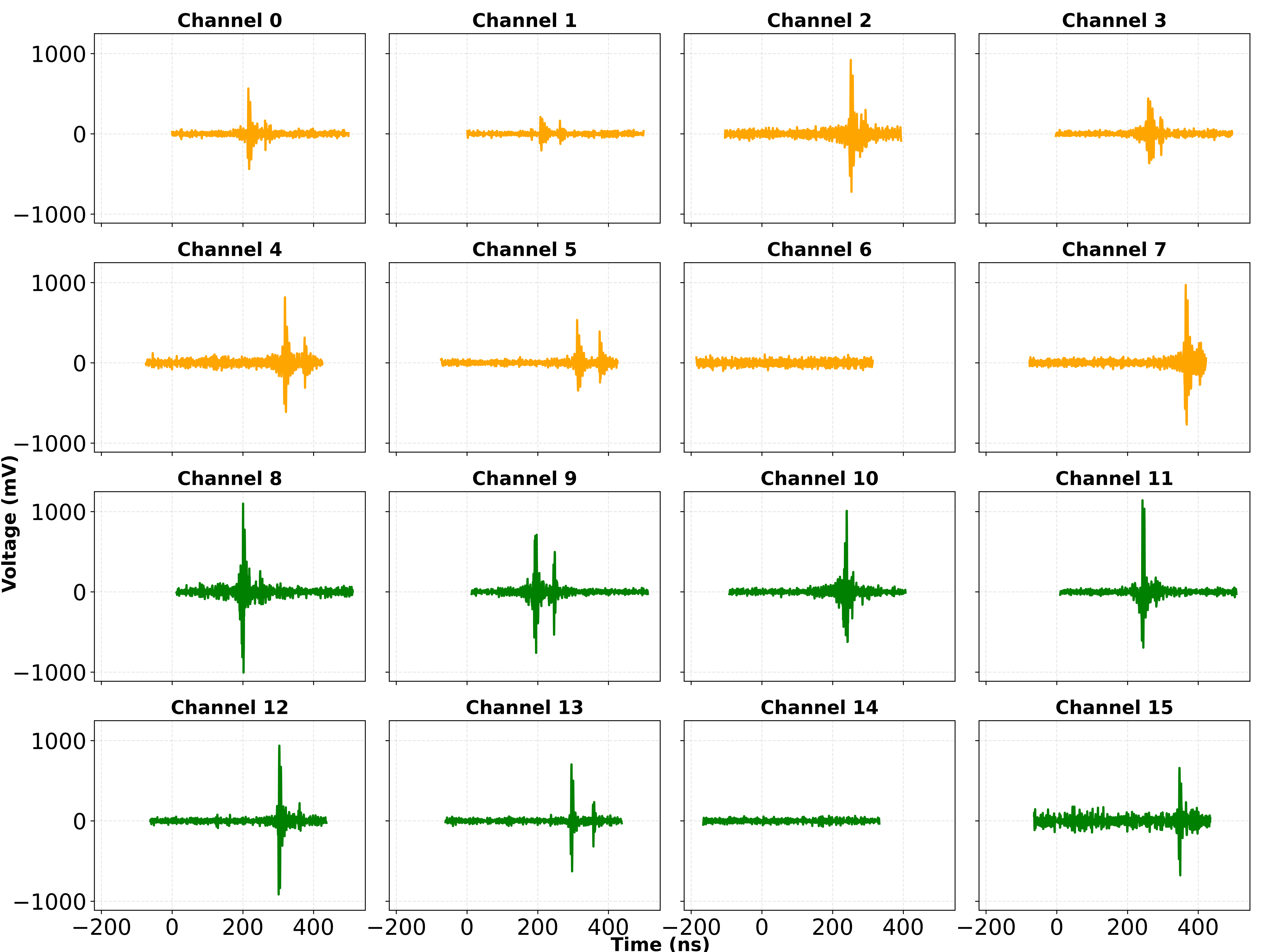}
    \caption{Waveforms(voltage traces in the time domain) of the CR candidate event. Double pulses with variable time delays can be seen in the figure. The top two rows correspond to the VPol channels, while the bottom two rows show the waveforms in the HPol channels.}
    \label{fig:event_wf}   
\end{figure}

After de-dispersing the measured effect of the ARA signal chain, we obtain the waveform shown in Figure~\ref{fig:event_wf}.
To perform independent vertex reconstruction for the two signals, the waveforms were split into first and second pulses based on their calculated hit times. The vertex reconstruction was conducted using the standard ARA interferometric reconstruction algorithm and cross-validated with the signal hit time-based tool, AraVertex~[\cite{Interferometry, AraVtx}]. The reconstructed vertex positions are expressed in the ARA station-centric coordinate system (ASC), as shown in Figure~\ref{fig:sim_event_rec}.
The ASC is a Cartesian reference frame, where the origin for Station A2 is defined at the station centroid depth of 179.9 m within the ice, and with the x-axis aligned with the direction of ice flow at the South Pole. The zenith angle ($\theta$) is measured relative to the vertical (z-axis), with $\theta=0^\circ$ corresponding to signals arriving from directly above the station. The azimuth angle ($\phi$) is measured counter-clockwise with respect to the x-axis in the horizontal plane. 

The time delays between the two pulses are not uniform across all channels, indicating the presence of two distinct emission sources rather than a single double-pulsing source. The time delay between the two pulses depends on the propagation paths and the media traversed by the signals from their respective sources to the receiving antennas. For a CR event, this time difference is influenced by the shower geometry, the antenna positions, the location of the shower impact point at the air–ice interface, and the densities of air and ice. A comparison between the measured time delays and those expected from geomagnetic and Askaryan emissions in a CR air shower is shown in  Figure~\ref{fig:dt}.

At the South Pole, the Earth's magnetic field is oriented almost vertically. Consequently, the geomagnetic emission produced by the Lorentz force acting on CR-induced air showers is expected to be predominantly horizontally polarized. In contrast, the Askaryan emission does not exhibit a specific polarization preference, as its VPol and HPol components contribute almost equally. Therefore, the power ratio defined as ($ \frac{\sum_{ch=0}^{16} A_{HPol}}{\sum_{ch=0}^{16} A_{VPol}}$)is expected to be higher for geomagnetic emission compared to the Askaryan emission. The signal amplitude $A$ for each channel is calculated using the relation: 
\begin{equation}
    A = \sqrt{A^2_{env}-V_{rms}^2}
\end{equation}
where $A_{env}$ is the amplitude of the Hilbert envelope of the pulse, and $V_{rms}$ represents the root-mean-squared voltage in the non-signal region. The amplitude is then averaged over all channels that show detectable peaks. For the CR candidate event, the ratio of the first-pulse power ratio to the second-pulse power ratio is 1.34.
\section{Simulations}\label{sim}
This analysis is supported by simulations performed using a recently developed Monte Carlo–based framework designed to model CR signals in in-ice detectors. This framework, called FAERIE (Framework for the Simulation of Air Shower Emission of Radio for In-Ice Experiments)~[\cite{faerie}], integrates CORSIKA 7.7500~[\cite{corsika}], a modified version of CoREAS, and GEANT4 10.5~[\cite{GEANT4}]. FAERIE includes a depth-dependent refractive index for both ice and air. The modified CoREAS module uses ray tracing to propagate the electric fields calculated in the endpoint formalism~[\cite{endpoint}]. These electric field traces are then processed through the ARA detector simulation (AraSim), which applies the detector response and produces time-domain voltage traces. An ARA-calibrated single exponential index of refraction model~[\cite{kh_ana}] for Antarctic ice is used in these simulations. 
\section{Results}\label{results}
\begin{figure}
    \centering
    \includegraphics[width=\linewidth]{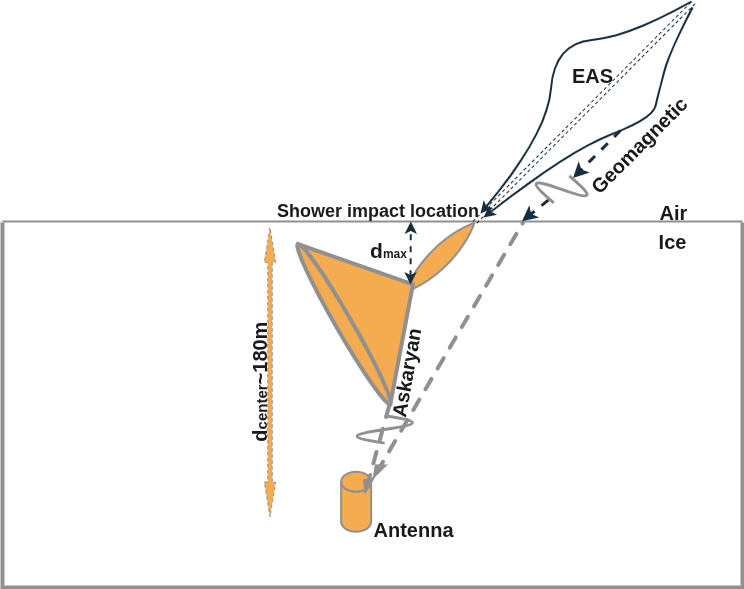}
    \caption{The predicted event geometry for the CR candidate event, where the in-ice antennas at the A2 center detect both geomagnetic emission produced in air and Askaryan emission generated within the ice.}
    \label{fig:ev_top}
\end{figure}

\begin{figure}
    \centering
    \includegraphics[width=\linewidth]{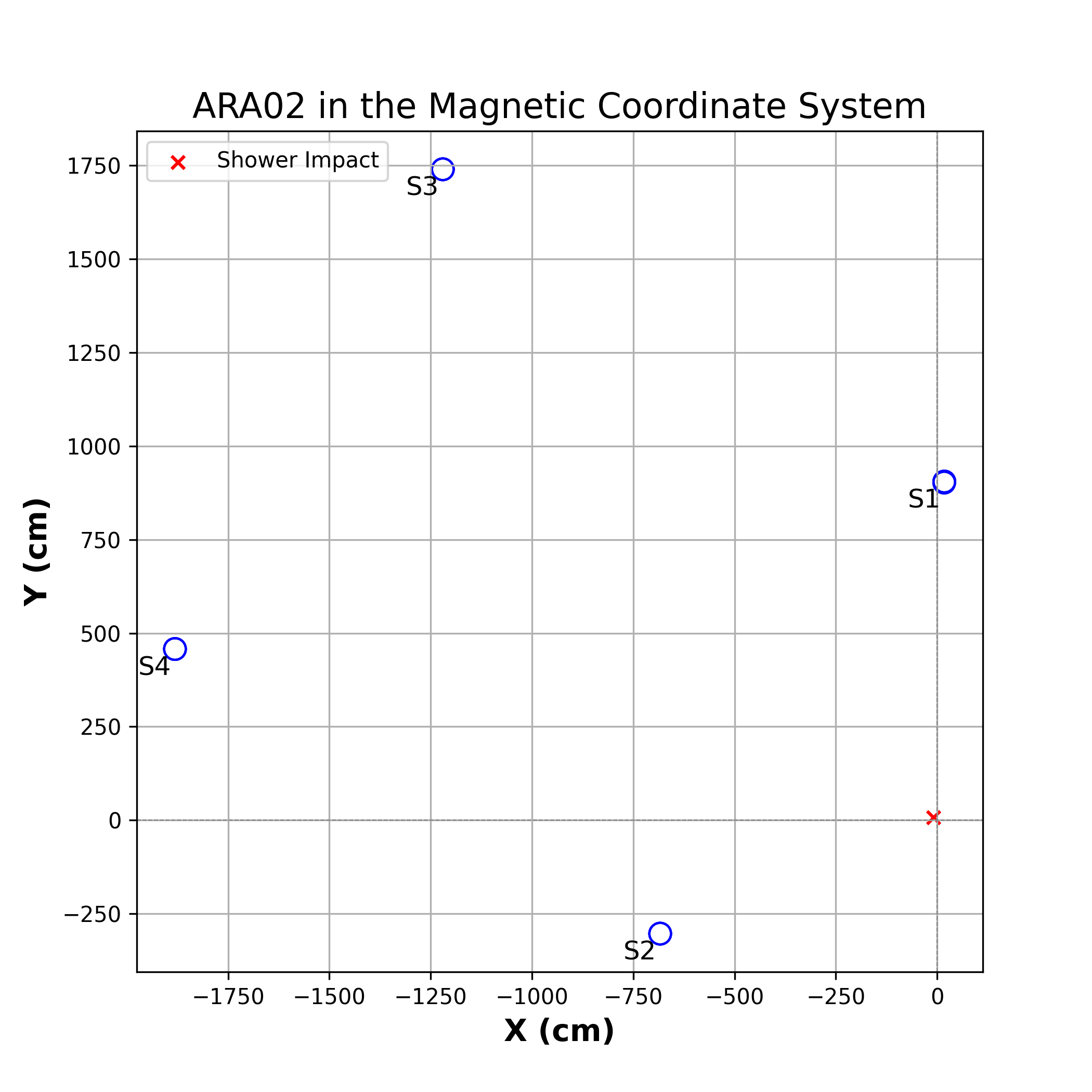}
    \caption{The four strings (S1 to S4) at the A2 station and their distances from the point where the shower core hits the ice (impact distance) are provided in a magnetic coordinate system. The simulated signals will be detected by antennas located along these strings at the specified x-y coordinates.}
    \label{fig:a2_geom}
\end{figure}

\begin{figure}
    \centering
    \includegraphics[width=\linewidth]{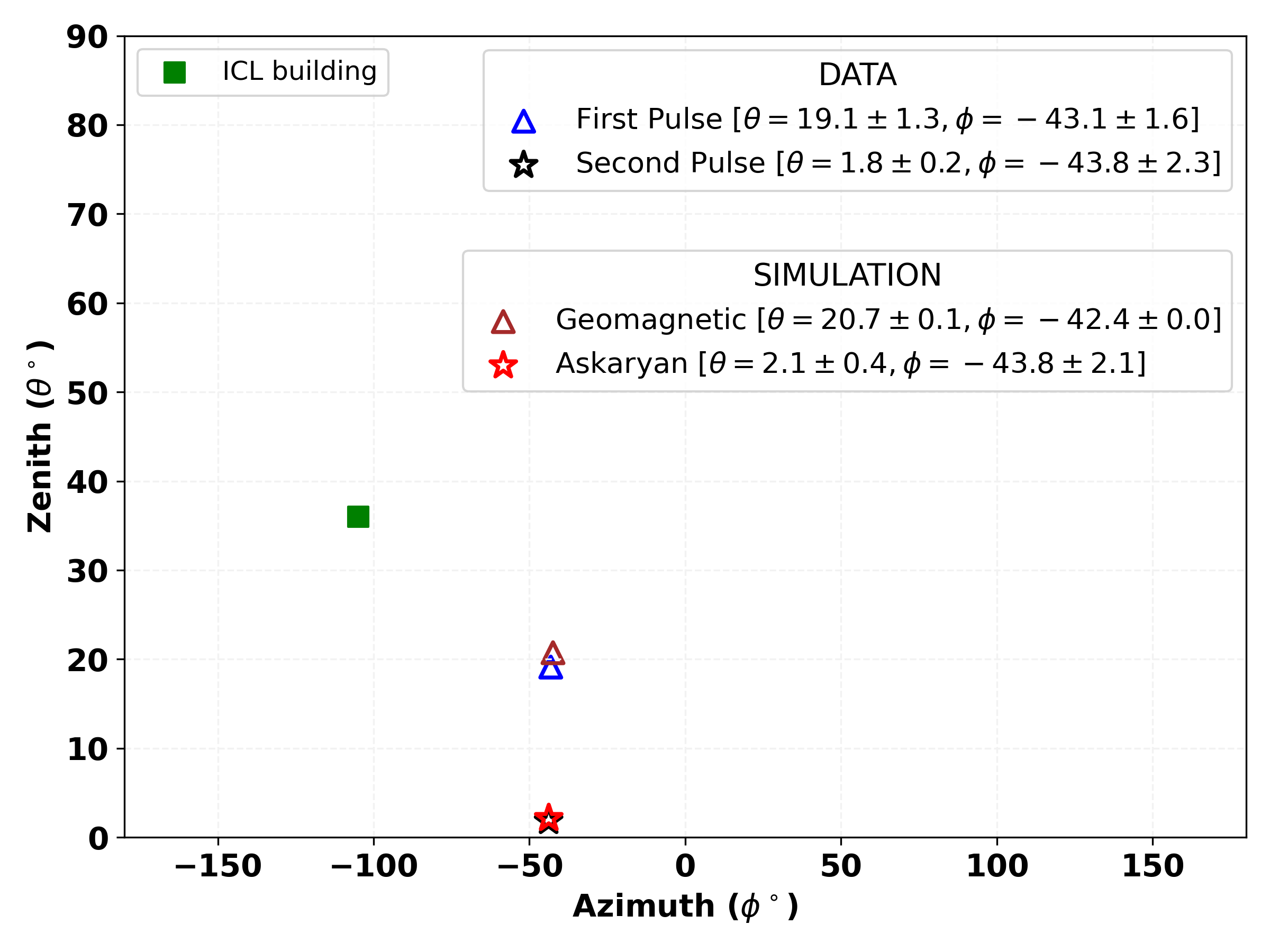}
    \caption{Vertex reconstruction directions for the first and second pulses are compared, respectively, with geomagnetic and Askaryan emissions recorded by the A2 station. Calculations are performed in the ASC coordinate system, with the origin at the center of the A2 station. The green box indicates the direction of the IceCube Lab (ICL), a known anthropogenic source at the South Pole. The errors are estimated from the ten simulated events.}
    \label{fig:sim_event_rec}
\end{figure}

\begin{figure}[htb!]
    \centering
    \includegraphics[width=\linewidth]{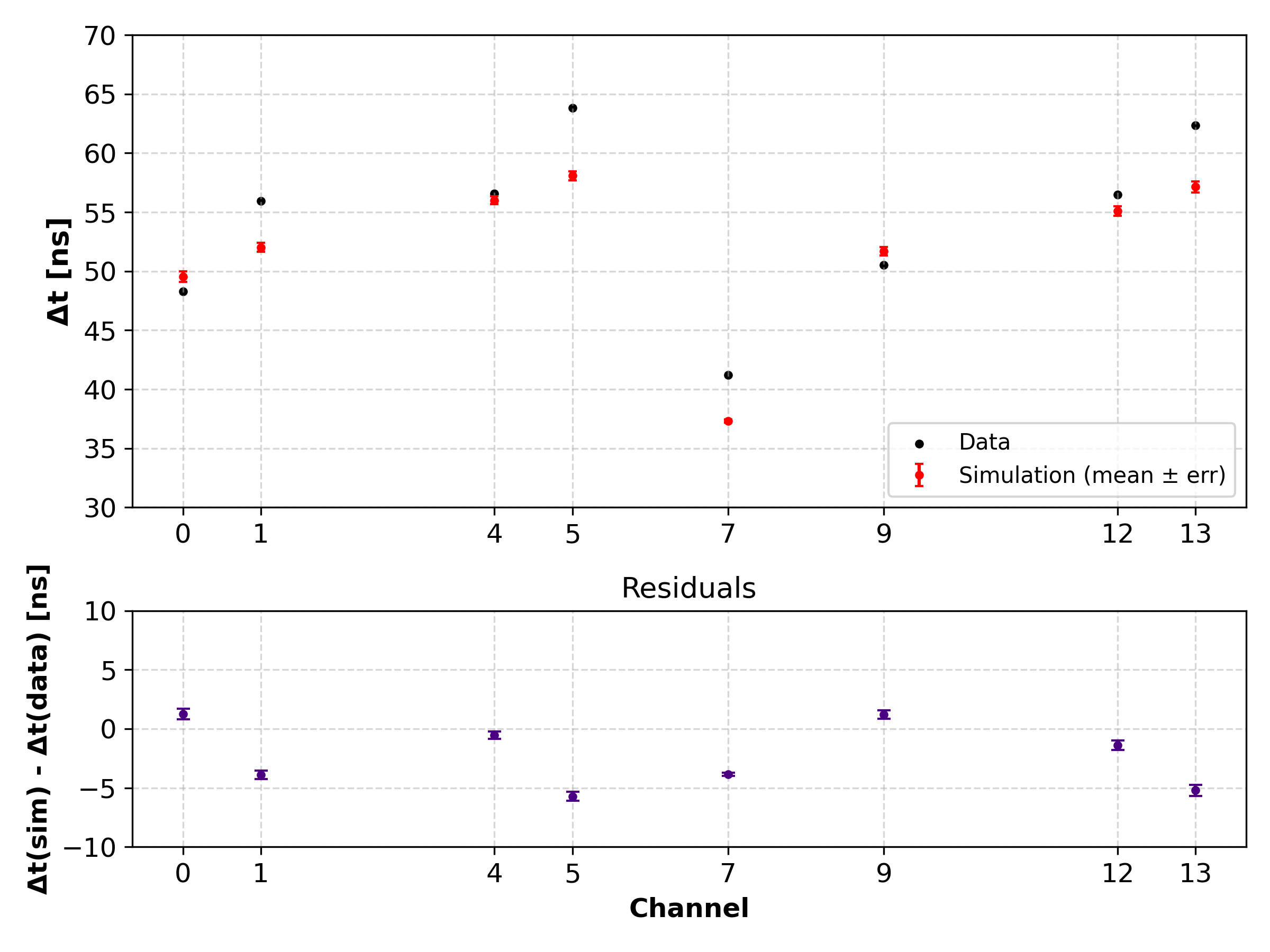}
    \caption{Channel-to-channel time delays, defined as the difference in arrival times between the two pulses, where the arrival time is determined by the peak of the pulse in the Hilbert envelope of the waveform. Errors are based on ten simulated events.}
    \label{fig:dt}
\end{figure}

To define the geometry, we must estimate the primary shower direction $(\theta_s, \varphi_s)$ and lateral impact distance $(d_s)$ relative to $(x,y)=(0,0)$ on the surface. These shower parameters are related to the reconstructed vertex directions $(\theta_{\mathrm{rec}}, \phi_{\mathrm{rec}})$ of the first and second pulses as:
\begin{subequations}
\begin{align}
    \theta_s &= \sin^{-1}\left(\frac{n_{\text{ice}} \cdot \sin(\theta_{\text{rec, 1st pulse}})}{n_{\text{air}}}\right) \tag{1} \\
    \varphi_s &= 180^\circ - \phi_{\text{rec}} \tag{2} \\
    d_s &= (d_{\text{center}} - d_{\text{max}}) \cdot \tan(\theta_{\text{rec, 2nd pulse}}) + d_{\text{max}} \cdot \tan(\theta_{\text{rec, 2nd pulse}} + \theta_c) \tag{3}
\end{align}
\end{subequations}
The station center depth ($d_{\mathrm{center}}$) for A2 is at $179.9~\mathrm{m}$, and the shower maximum in ice is taken to occur at a depth of $6~\mathrm{m}$, as determined from simulated CR showers. The bulk refractive index of ice is assumed to be $n_{\mathrm{ice}} = 1.78$, and the Cherenkov angle ($\theta_c$) at the shower maximum is set to $43^\circ$. The implied event geometry, constrained by these parameters, is shown in  Figure~\ref{fig:ev_top}, and the four string coordinates with the impact location of the shower given in the magnetic coordinate system are presented in Figure~\ref{fig:a2_geom}.
To probe shower-to-shower fluctuation effects, ten CR-induced air shower events were simulated according to the predicted event geometry for the A2 station. The simulated data were analyzed to reconstruct the vertex direction for both geomagnetic and Askaryan emissions in the ASC coordinate system. The reconstructed vertices were compared with those obtained from the first and second pulses, respectively, and the results are shown in  Figure~\ref{fig:sim_event_rec}.
Since the geomagnetic and Askaryan emissions originate from distinct locations and propagate along different paths for each receiver antenna, the arrival time differences between the two signals are not expected to be uniform across channels. Figure~\ref{fig:dt} presents the simulated time delays between the geomagnetic and Askaryan emissions compared with the measured data, showing very close agreement (albeit with a slight offset from zero), with residuals of order a few ns.
To estimate the polarization characteristics, the power ratio was calculated from the AraSim waveforms. The preliminary simulated value for the geomagnetic relative to the Askaryan emission ratio is $1.74\pm 0.47$, with the error corresponding to the rms of our 10 simulated events. The effect of (known) density variations in the upper ice layers where the shower develops has not yet been evaluated.
\section{Conclusion and Outlook}
This study proposes a topology for the A2 CR candidate event. The reconstructed vertex directions, obtained from simulated geomagnetic and Askaryan radio emissions of a CR–induced air shower, show good agreement with the measured vertex directions corresponding to the first and second pulses in the data, with residuals smaller than $2^\circ$. The measured time delays between the two pulses also closely match the simulated delays, with residuals less than 5 ns between the geomagnetic and Askaryan components across different channels.\\
The results for the power ratio of the geomagnetic to Askaryan components further demonstrate consistent alignment between simulation and observation, despite significant event-to-event amplitude fluctuations.\\
Ongoing work aims to further tune the relative power contributions of the two emission mechanisms and to estimate the primary CR energy associated with this event. We emphasize that this study is still in progress. Therefore, the results presented here should be regarded as preliminary.
\setlength{\itemsep}{0pt}
\setlength{\parsep}{0pt}
\setlength{\parskip}{0pt}

\clearpage
\onecolumn
\section*{Acknowledgements}
\noindent
The ARA Collaboration is grateful to support from the National Science Foundation through Award 2013134.
The ARA Collaboration designed, constructed, and now operates the ARA detectors. 
We would like to thank IceCube, and specifically the winterovers for the support in operating the detector. 
Data processing and calibration, Monte Carlo simulations of the detector and of theoretical models and data analyses were performed by a large number
of collaboration members, who also discussed and approved the scientific results presented here. 
We are thankful to Antarctic Support Contractor staff, a Leidos unit for field support and enabling our work on the harshest continent. 
We thank the National Science Foundation (NSF) Office of Polar Programs and Physics Division for funding support. 
We further thank the Taiwan National Science Councils Vanguard Program NSC 92-2628-M-002-09 and the Belgian F.R.S.-FNRS and FWO.
K.~Hughes thanks the NSF for support through the Graduate Research Fellowship Program Award DGE-1746045. 
A.~Connolly thanks the NSF for Award 1806923 and 2209588, and also acknowledges the Ohio Supercomputer Center. 
S.~A.~Wissel thanks the NSF for support through CAREER Award 2033500.
A.~Vieregg, C.~Deaconu, N.~Alden, and P.~Windischhofer thank the NSF for Award 2411662 and the Research Computing Center at the University of Chicago
for computing resources.
R.~Nichol thanks the Leverhulme Trust for their support. 
K.D.~de~Vries is supported by European Research Council under the European Unions Horizon research and innovation program (grant agreement 763 No 805486). 
D.~Besson, I.~Kravchenko, and D.~Seckel thank the NSF for support through the IceCube EPSCoR Initiative (Award ID 2019597). 
M.S.~Muzio thanks the NSF for support through the MPS-Ascend Postdoctoral Fellowship under Award 2138121. 
A.~Bishop thanks the Belgian American Education Foundation for their Graduate Fellowship support.

\section*{Full Author List: ARA Collaboration (November 07, 2025)}

\noindent
N.~Alden\textsuperscript{1}, 
S.~Ali\textsuperscript{2}, 
P.~Allison\textsuperscript{3}, 
S.~Archambault\textsuperscript{4}, 
J.J.~Beatty\textsuperscript{3}, 
D.Z.~Besson\textsuperscript{2}, 
A.~Bishop\textsuperscript{5}, 
P.~Chen\textsuperscript{6}, 
Y.C.~Chen\textsuperscript{6}, 
Y.-C.~Chen\textsuperscript{6}, 
S.~Chiche\textsuperscript{7}, 
B.A.~Clark\textsuperscript{8}, 
A.~Connolly\textsuperscript{3}, 
K.~Couberly\textsuperscript{2}, 
L.~Cremonesi\textsuperscript{9}, 
A.~Cummings\textsuperscript{10,11,12}, 
P.~Dasgupta\textsuperscript{3}, 
R.~Debolt\textsuperscript{3}, 
S.~de~Kockere\textsuperscript{13}, 
K.D.~de~Vries\textsuperscript{13}, 
C.~Deaconu\textsuperscript{1}, 
M.A.~DuVernois\textsuperscript{5}, 
J.~Flaherty\textsuperscript{3}, 
E.~Friedman\textsuperscript{8}, 
R.~Gaior\textsuperscript{4}, 
P.~Giri\textsuperscript{14}, 
J.~Hanson\textsuperscript{15}, 
N.~Harty\textsuperscript{16}, 
K.D.~Hoffman\textsuperscript{8}, 
M.-H.~Huang\textsuperscript{6,17}, 
K.~Hughes\textsuperscript{3}, 
A.~Ishihara\textsuperscript{4}, 
A.~Karle\textsuperscript{5}, 
J.L.~Kelley\textsuperscript{5}, 
K.-C.~Kim\textsuperscript{8}, 
M.-C.~Kim\textsuperscript{4}, 
I.~Kravchenko\textsuperscript{14}, 
R.~Krebs\textsuperscript{10,11}, 
C.Y.~Kuo\textsuperscript{6}, 
K.~Kurusu\textsuperscript{4}, 
U.A.~Latif\textsuperscript{13}, 
C.H.~Liu\textsuperscript{14}, 
T.C.~Liu\textsuperscript{6,18}, 
W.~Luszczak\textsuperscript{3}, 
A.~Machtay\textsuperscript{3}, 
K.~Mase\textsuperscript{4}, 
M.S.~Muzio\textsuperscript{5,10,11,12}, 
J.~Nam\textsuperscript{6}, 
R.J.~Nichol\textsuperscript{9}, 
A.~Novikov\textsuperscript{16}, 
A.~Nozdrina\textsuperscript{3}, 
E.~Oberla\textsuperscript{1}, 
C.W.~Pai\textsuperscript{6}, 
Y.~Pan\textsuperscript{16}, 
C.~Pfendner\textsuperscript{19}, 
N.~Punsuebsay\textsuperscript{16}, 
J.~Roth\textsuperscript{16}, 
A.~Salcedo-Gomez\textsuperscript{3}, 
D.~Seckel\textsuperscript{16}, 
M.F.H.~Seikh\textsuperscript{2}, 
Y.-S.~Shiao\textsuperscript{6,20}, 
J.~Stethem\textsuperscript{3}, 
S.C.~Su\textsuperscript{6}, 
S.~Toscano\textsuperscript{7}, 
J.~Torres\textsuperscript{3}, 
J.~Touart\textsuperscript{8}, 
N.~van~Eijndhoven\textsuperscript{13}, 
A.~Vieregg\textsuperscript{1}, 
M.~Vilarino~Fostier\textsuperscript{7}, 
M.-Z.~Wang\textsuperscript{6}, 
S.-H.~Wang\textsuperscript{6}, 
P.~Windischhofer\textsuperscript{1}, 
S.A.~Wissel\textsuperscript{10,11,12}, 
C.~Xie\textsuperscript{9}, 
S.~Yoshida\textsuperscript{4}, 
R.~Young\textsuperscript{2}
\\
\\
\textsuperscript{1} Dept. of Physics, Dept. of Astronomy and Astrophysics, Enrico Fermi Institute, Kavli Institute for Cosmological Physics, University of Chicago, Chicago, IL 60637\\
\textsuperscript{2} Dept. of Physics and Astronomy, University of Kansas, Lawrence, KS 66045\\
\textsuperscript{3} Dept. of Physics, Center for Cosmology and AstroParticle Physics, The Ohio State University, Columbus, OH 43210\\
\textsuperscript{4} Dept. of Physics, Chiba University, Chiba, Japan\\
\textsuperscript{5} Dept. of Physics, University of Wisconsin-Madison, Madison,  WI 53706\\
\textsuperscript{6} Dept. of Physics, Grad. Inst. of Astrophys., Leung Center for Cosmology and Particle Astrophysics, National Taiwan University, Taipei, Taiwan\\
\textsuperscript{7} Universite Libre de Bruxelles, Science Faculty CP230, B-1050 Brussels, Belgium\\
\textsuperscript{8} Dept. of Physics, University of Maryland, College Park, MD 20742\\
\textsuperscript{9} Dept. of Physics and Astronomy, University College London, London, United Kingdom\\
\textsuperscript{10} Center for Multi-Messenger Astrophysics, Institute for Gravitation and the Cosmos, Pennsylvania State University, University Park, PA 16802\\
\textsuperscript{11} Dept. of Physics, Pennsylvania State University, University Park, PA 16802\\
\textsuperscript{12} Dept. of Astronomy and Astrophysics, Pennsylvania State University, University Park, PA 16802\\
\textsuperscript{13} Vrije Universiteit Brussel, Brussels, Belgium\\
\textsuperscript{14} Dept. of Physics and Astronomy, University of Nebraska, Lincoln, Nebraska 68588\\
\textsuperscript{15} Dept. Physics and Astronomy, Whittier College, Whittier, CA 90602\\
\textsuperscript{16} Dept. of Physics, University of Delaware, Newark, DE 19716\\
\textsuperscript{17} Dept. of Energy Engineering, National United University, Miaoli, Taiwan\\
\textsuperscript{18} Dept. of Applied Physics, National Pingtung University, Pingtung City, Pingtung County 900393, Taiwan\\
\textsuperscript{19} Dept. of Physics and Astronomy, Denison University, Granville, Ohio 43023\\
\textsuperscript{20} National Nano Device Laboratories, Hsinchu 300, Taiwan\\

\end{document}